\RequirePackage{fix-cm}
\documentclass[11pt,a4paper]{article}
\usepackage{graphicx}

\usepackage{amsmath,latexsym,psfrag,epsf,epsfig,amssymb,rotating,dsfont}
\usepackage{natbib}

\usepackage{geometry} 
\begin{document}

\title{Efficient computational strategies for doubly intractable problems with applications to Bayesian social networks}
\author{Alberto Caimo\\Faculty of Economics\\ University of Lugano\\ Switzerland
  \and Antonietta Mira\\Institute of Finance\\ University of Lugano\\ Switzerland}

\maketitle

\begin{abstract}
Powerful ideas recently appeared in the literature are adjusted and combined to design improved samplers for 
doubly intractable target distributions with a focus on
Bayesian exponential random graph models.
Different forms of adaptive Metropolis-Hastings proposals (vertical, horizontal and rectangular)  are tested and merged with 
the delayed rejection (DR) strategy with the aim of reducing the variance of the resulting Markov chain Monte Carlo estimators for a given computational time. 
The DR is modified in order to integrate it within the approximate exchange algorithm (AEA) to avoid the computation of intractable normalising constant that appears in exponential random graph models. This gives rise to the AEA+DR: a new methodology to sample doubly intractable distributions that dominates the AEA  in the Peskun
ordering \citep{pes73} leading to MCMC estimators with a smaller asymptotic variance.
The  {\sf Bergm} package for {\sf R} \citep{cai:fri14} has been updated to incorporate the AEA+DR thus
including the possibility of adding a higher stage proposals and different forms of adaptation. 

\end{abstract}

\section{Introduction}
\label{sec:intro}

In this paper we combine the 
approximate exchange algorithm (AEA) proposed in \cite{cai:fri11}, which has been proven to be particularly efficient in estimating exponential random graph models (ERGMs), with the delayed rejection (DR)  introduced in \cite{tie:mir99}, a strategy to reduce the asymptotic variance of the resulting MCMC estimators. In particular we focus on the adaptive direction sampling approximate exchange algorithm
(ADS-AEA) which is based on the idea of running, in parallel, multiple chains that, at each fixed simulation time, interact with each other to allow the construction of a distribution that selects the proposal direction of the candidate move by picking at random a pair of chains.

We also suggest an alternative to ADS-AEA based on an adaptive random walk proposal distribution. 
Three different adaptation strategies will be studied to design a good proposal variance-covariance matrix: the first one is based on the past history of each single chain 
(vertical adaptation); the second  is based on the current population of all chains at the given simulation time 
(horizontal adaptation), and finally global adaptation takes into account the past history of all chains 
(rectangular adaptation).

The three ingredients (ADS, DR and Adaptive proposal) are combined in various ways 
 and compared to obtain the most effective strategy. Optimality is measure by the effective sample size (ESS) and the performance (defined as ESS per simulation time) 
 and the focus is on estimating ERGMs.
 
 The novel methodological contribution consists in the new second (and higher stage) acceptance probability of the approximate exchange algorithm with delayed rejection (AEA+DR) that does not require the calculation of the likelihood normalising constant and can thus be used to generate a Markov chain having a general doubly intractable posterior target as its stationary distribution. The DR strategy leads, by construction, to MCMC estimators that have a smaller asymptotic variance. Indeed the AEA-DR dominates, in the Peskun sense \citep{pes73}, the regular AEA.

\section{Exponential Random Graph Models}
\label{sec:ERGMs}
Exponential random graph models (see \cite{rob:sni:wan:han:pat07} for a recent review) assume that the topological structure  in an observed network $y$ can be explained by the relative prevalence of a set of overlapping sub-graph configurations $s(y)$ also called graph or network statistics.

Each network statistic has an associated unknown parameter. 
A positive value for a certain parameter $\theta^{(i)}$ indicates that the edges involved in the formation of the corresponding network statistic $s_i(y)$ are more likely to be observed relative to edges that are not involved in the formation of that network statistic, and vice versa.

Network statistics and parameters are at the core of ERGMs and the challenge is to estimate the parameters for each statistic such that the model is a good fit for the given data. From a statistical point of view, networks are relational data represented as mathematical graphs. A graph consists of a set of $n$ nodes and a set of $m$ ties which define a relationship between pairs of nodes called dyads. The connectivity pattern of a graph can be described by an $n \times n$ adjacency matrix $y$ encoding the presence or absence of a tie between node $i$ and $j$: $y_{ij} = 1$ if the dyad $(i,j)$ is connected, $y_{ij} = 0$ otherwise. 
The likelihood of an ERGM represents the probability distribution of a random network graph and can be expressed as:
\begin{equation}
p(y|\theta) =  \frac{q(y|\theta)}{z(\theta)} = \frac{\exp\{s(y)^T \theta \}}{z(\theta)}
\label{eqn:ergm}
\end{equation}
where $q(y|\theta)$ is the unnormalised likelihood.
This equation states that the probability of observing a given network graph $y$ is equal to the exponent of the observed graph statistics $s(y)$ multiplied by parameter vector $\theta$ divided by a normalising constant term $z(\theta)$. The latter is calculated over the sum of all possible graphs on $n$ nodes and it is therefore extremely difficult to evaluate for all but trivially small graphs since this sum involves $2^{\binom{n}{2}}$ terms (for undirected graphs).
The intractable normalising constant makes inference difficult for both frequentist and Bayesian approaches. This problem does not only occur in ERGMs, but in many other statistical models including, for example, the autologistic model \citep{bes74} in spatial statistics. Given the similarities among these models  from a computational tractability point of view, we envisage that the MCMC simulation strategies  proposed in this paper are amenable of successful application in these other contexts as well.

\section{Bayesian Methods for ERGMs}
\label{sec:Bayes}
Bayesian methods are becoming increasingly popular as techniques for modelling social networks. In the ERGM context recent works on using the Bayesian approach for inferring ERGMs have been proposed by \cite{kos:rob:pat10} and \cite{cai:fri11,cai:fri13}. 

Following the Bayesian paradigm, a prior distribution is assigned to $\theta$. The posterior distribution of $\theta$ given the data $y$ is:
\begin{equation}
p(\theta | y ) = \frac{p(y|\theta) p(\theta)}{p(y)}.
\end{equation}
Direct evaluation of $p(\theta|y)$ requires the calculation of both the likelihood $p(y|\theta)$ and the marginal likelihood $p(y)$ which are typically intractable. For this reason posterior parameter estimation for ERGMs has been termed a doubly-intractable problem.

\subsection{Exchange Algorithm}
\label{sec:exchange}
Markov chain Monte Carlo (MCMC) algorithms \citep{tie94} are general simulation methods for sampling from posterior distributions and computing posterior quantities of interest. 
The most widely used MCMC sampler is the Metropolis-Hastings (MH) that, under easy to verify regularity conditions, constructs an ergodic Markov chain having the posterior $p(\theta|y) \propto p(y|\theta) p(\theta)$ as its unique stationary and limiting distribution.

A na\"{i}ve MH update,  proposing to move from the current state $\theta$ to $\theta_1$, would require calculation of the following acceptance probability at each sweep of the algorithm:
\begin{equation} 
\alpha(\theta, \theta_1) = 1 \wedge \frac{q(y|\theta_1)p(\theta_1) h(\theta|\theta_1)}{q(y|\theta) p(\theta) h(\theta_1|\theta)} \times \frac{z(\theta)}{z(\theta_1)} 
\label{eqn:naive}
\end{equation}
where $q(\cdot)$ represents the unnormalised likelihood and $h(\cdot)$ is a proposal distribution used to generate the candidate move $\theta_1$.
For doubly intractable target distributions, 
the ratio in (\ref{eqn:naive}) is unworkable due to the presence of the normalising constants $z(\theta)$ and $z(\theta_1)$ 
(note that, on the other hand, the marginal likelihood cancels and thus one source of intractability is resolved).

A special case of the MH algorithm is the random-walk MH, where the proposal (typically a Gaussian distribution) is centred at the current position of the Markov chain and thus
$\theta_1 = \theta + \sigma \, \epsilon$ where $\epsilon$ is, usually, a standard Gaussian displacement. Since this proposal $h$ is symmetric i.e. $h(\theta|\theta_1) = h(\theta_1|\theta)$, it cancels in the acceptance ratio.
A typical difficulty in the MH algorithm is the proper tuning of the proposal distribution that translates, for the random-walk MH in the choice of the tuning parameter $\sigma$.
Off-line tuning aiming at achieving the optimal (in some high dimensional context) acceptance rate of approximately 0.234 \citep{rob:gel:gil97,rob:ros98,rob:ros01} is possible but time consuming.
A recent  better alternative is adaptive on-line design of the proposal: when tuning the proposal at simulation time $t$ the whole past history of the chain can be taken into account. 
Different forms of adaptations are possible but
since these adaptive strategies destroy the Markovian properties of the sampler, careful rules should be followed in on-line adaptation procedures \citep{and:atc06,and:mou06,rob:ros07,atc:ros05}.
Another possibility is to run in parallel more Markov chains, all having the same target distribution, and when designing the proposal for one of the chains learn from the current position of the other ones. This strategy does not destroy the Markovian property of the chain being updated and thus it is easier to adopt and gives more freedom in designing adaption strategies, but has additional computational costs. This is the reason why, when comparing alternative adaptation strategies, simulation time should be taken into account.

To get around the issue related to the intractability of the likelihood and thus of the MH acceptance probability,
\cite{mur:gha:mac06} proposed to estimate $\frac{z(\theta)}{z(\theta_1)}$ directly, by considering the following augmented distribution:
\begin{equation}
p(\theta_1,y_1,\theta|y) \propto p(y|\theta)p(\theta) h(\theta_1|\theta) \times p(y_1|\theta_1) 
\end{equation}
where $y_1$ are auxiliary data generated from the distribution $p(\cdot|\theta_1)$ which is the same distribution from which the observed data $y$ are assumed to have been sampled from. Notice that the original target is a proper marginal  of the augmented distribution thus, running a Markov chain on the augmented state space and marginalising over $\theta$, returns an ergodic sample from the proper posterior of interest.

Using this augmented distribution has the advantage that the acceptance probability in  (\ref{eqn:naive}) can be written as:
\begin{equation}
\label{eqn:acc_prob}
1 \wedge \frac{q(y|\theta_1)p(\theta_1) h(\theta|\theta_1)}
{q(y|\theta)p(\theta) h(\theta_1|\theta)} 
\times \frac{q(y_1|\theta)}{q(y_1|\theta_1)}
\times \frac{z(\theta)}{z(\theta_1)}
\times \frac{z(\theta_1)}{z(\theta)}.
\end{equation}
All intractable normalising constants cancel above and below the ratio making the acceptance probability \eqref{eqn:acc_prob} 
of the Metropolis-Hastings algorithm on the enlarged state space, computable.

\subsection{Adaptive Direction Sampling Approximate Exchange Algorithm (ADS-AEA)}
\label{sec:exchange-ads}
The exchange algorithm of \cite{mur:gha:mac06} requires exact simulation of new data $y_1$ from the likelihood $p(\cdot|\theta_1)$. However in the ERGM context, and more generally in Gibbs random fields, exact sampling from the likelihood is difficult. \cite{cai:fri11} proposed to approximate the exact simulation of $y_1$ from $p(\cdot|\theta_1)$ using MCMC. A theoretical justification for the validity of this approach has been given by \cite{eve12}.

In order to improve mixing \cite{cai:fri11}  use an adaptive direction sampling (ADS)
method \citep{gil:rob:geo94,rob:gil94} similar to that of \cite{ter:vru08}.
The approach consists in running in parallel 
a collection of  $H$  chains which interact with one another.
The ADS move, as illustrated in \cite{cai:fri11},
can be described as follows. Set a scalar value for $\gamma$ (ADS move factor), for each chain $h$: 
\begin{enumerate}
\item Sample two current states $\theta^{h_1}$ and $\theta^{h_2}$ without replacement from the population $\{1,\dots,H\} \setminus h$
\item Sample $\epsilon$ from a symmetric proposal distribution
\item Propose $\theta_1^h = \theta^h + \gamma \left( \theta^{h_1} - \theta^{h_2}\right) + \epsilon$
\item Sample $y_1$ from $p(\cdot|\theta_1^h)$
\item Accept the move from $\theta^h$ to $\theta^h_1$ with probability
\begin{equation}
\alpha (\theta^h, \theta^h_1)  = 1\wedge  \frac{q(y|\theta^h_1)\;p(\theta^h_1)\;q(y_1|\theta^h)}
                                                                         {q(y|\theta^h)\;p(\theta^h)\;q(y_1|\theta^h_1)}.
\end{equation}
\end{enumerate} 
Note that, since the ADS proposal distribution is symmetric, it does not appear in the acceptance probability.

\begin{figure}
\centering
\includegraphics[scale=1]{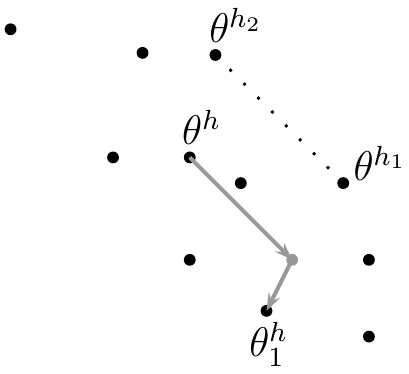}
\caption{The move of $\theta^h$ is generated from the difference $\theta^{h_1} - \theta^{h_2}$ plus a random term $\epsilon$.}
\label{fig:snooker1}
\end{figure}

\subsection{Florentine Marriage Network}
\label{sec:flo1}
\begin{figure}[htp]
\centering
\includegraphics[scale=.45]{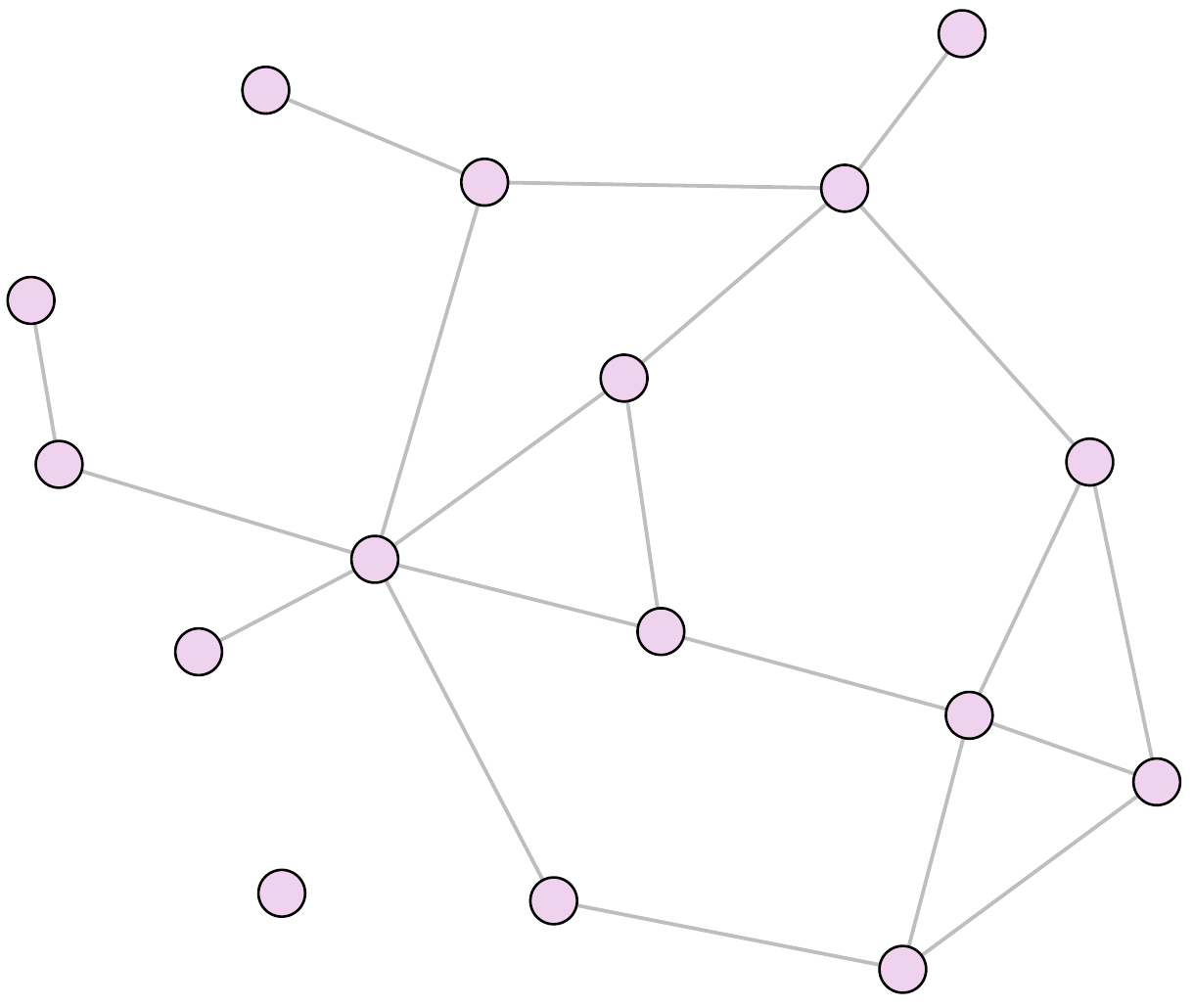}
\caption{Florentine marriage network graph.}
\label{fig:flo-graph}
\end{figure}

Let us consider, as a toy example, the 16-node Florentine marriage network data concerning the marriage relations between some Florentine families in around 1430 \citep{pad:ans93}. The network graph is displayed in Figure~\ref{fig:flo-graph}.

We propose to estimate the posterior distribution of the following 3-dimensional ERGM:
\begin{equation}
q(y|\theta) = \exp \left\lbrace \theta^{(1)} s_{1}(y) + \theta^{(2)} s_{2}(y) + \theta^{(3)}  s_{3}(y) \right\rbrace 
\label{eqn:model1}
\end{equation}
where
\begin{center}
\begin{tabular}{ll}
$s_{1}(y) = \sum_{i<j}y_{ij}$ & number of edges\\
$s_{2}(y) = \sum_{i<j<k}y_{ik}y_{jk}$ & number of 2-stars\\
$s_{3}(y) = \sum_{i<j<l<k}y_{ik}y_{jk}y_{lk}$ & number of 3-stars.
\end{tabular}
\end{center}
A vague multivariate Normal prior $p(\theta) \sim \mathcal{N}(0,100 I_d)$ is chosen, where $I_d$ is the identity matrix with dimensions equal to that of the model (the same prior setting will be used for all the examples in this paper).
The {\sf Bergm} package for {\sf R} \citep{cai:fri14} allows to carry out inference with the approximate exchange algorithm described above. 

We set the ADS move factor $\gamma = 0.8$ and  $\epsilon \sim \mathcal{N}(0,0.025 I_d)$. The auxiliary chain used to simulate auxiliary network data from the model consists of $50$ iterations and the main chain of $4,000$ iterations for each of the 6 chains of the MCMC population so that we have a total of $24,000$ main iterations. 
The tuning parameters were chosen so that the overall acceptance rate is around $21\%$. Table~\ref{tab:flo1} shows the posterior estimates and effective sample size (ESS) \citep{kas:car:gel98} which is calculated for each parameter $\theta^{(i)}$, $i = 1,\dots,d$:
\begin{equation*}
ESS(\theta^{(i)}) = \frac{S}{(1 + 2 \sum_k \rho_{k}(\theta^{(i)}))},
\end{equation*}
where $S$ is the number of posterior samples and $\rho_{k}(\cdot)$ is the autocorrelation at lag $k$. The infinite sum is often truncated at lag $k$ when $\rho_{k}(\theta^{(i)}) < 0.05$. 

The results indicate the tendency to a low number of edges as expressed by the edge parameter 
$\theta^{(1)}$ and null parameter values for $\theta^{(2)}$ and $\theta^{(3)}$. These estimates are consistent with the ones obtained using a frequentist approach \citep{hun:han:but:goo:mor08} as expected given the fairly vague prior.

\begin{table}
\caption{Florentine marriage network - Posterior parameter estimates and effective sample size (ESS).}
\label{tab:flo1}
\begin{tabular}{l|ccc}
\hline\noalign{\smallskip}
       & $\theta^{(1)}$ (edges) & $\theta^{(2)}$ (2-stars) & $\theta^{(3)}$ (3-stars) \\
\noalign{\smallskip}\hline\noalign{\smallskip}
Post. mean & -1.57 & 0.08 & -0.07 \\
Post. sd      & 1.93  &  0.71 & 0.34\\
\noalign{\smallskip}\hline\noalign{\smallskip}
ESS & 736 & 743 & 760 \\
\noalign{\smallskip}\hline
\end{tabular}
\end{table}

\begin{figure}
\includegraphics[scale=.7]{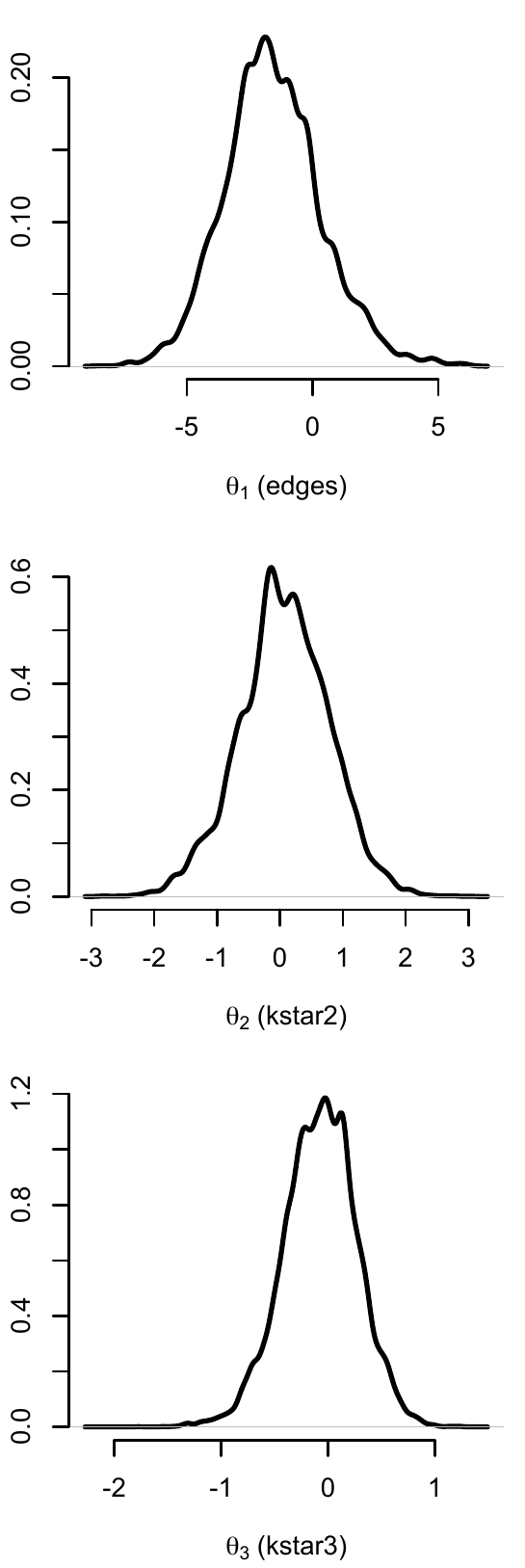} 
\includegraphics[scale=.7]{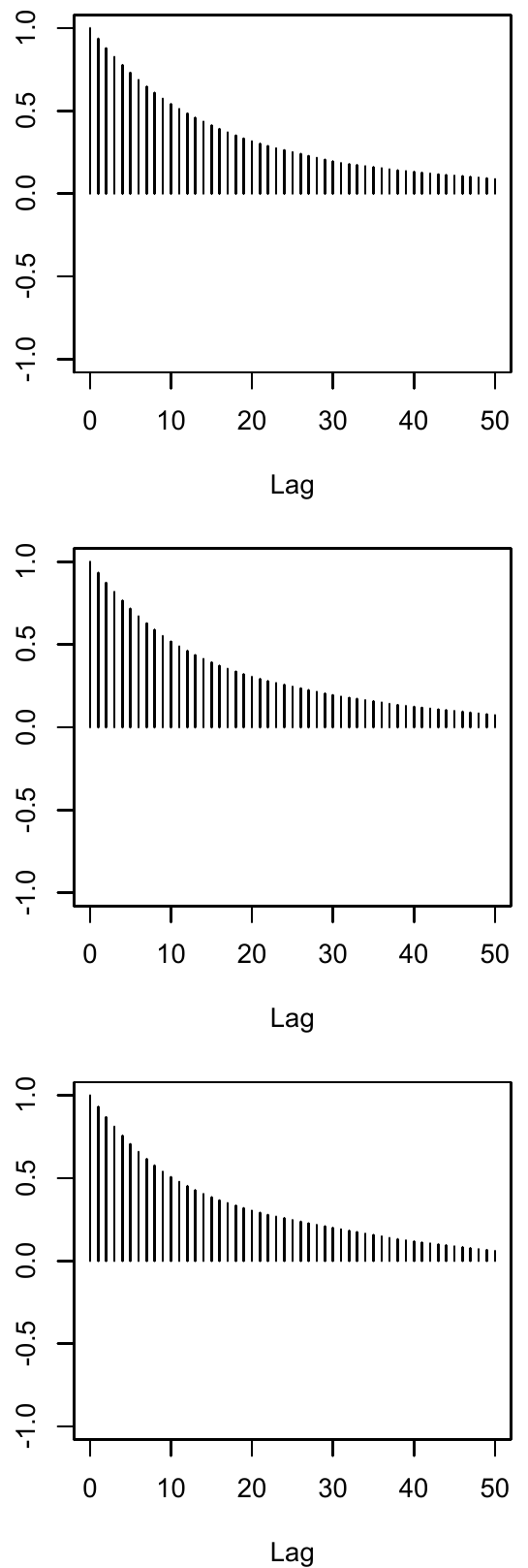} 
\caption{Florentine marriage network. MCMC diagnostics for the overall chain. The 2 plot columns are: estimated marginal posterior densities (left), and autocorrelation plots (right).}
\label{fig:flo0}
\end{figure}

\section{Delayed Rejection Strategy}
\label{sec:DR}
Delayed rejection (DR) is a modification of
the Me\-tro\-po\-lis-Hastings MCMC algorithm introduced in \cite{tie:mir99}
and generalized in \cite{gre:mir01,mir01}, aimed at improving
efficiency of the resulting MCMC estimators relative to asymptotic variance orderings introduced in 
\cite{pes73} and generalized by \cite{tie98,mir01a}. The basic idea
is that, upon rejection in a MH, instead of advancing simulation time and
retaining the same position of the Markov chain, a second stage move is proposed.  The
acceptance probability of the second stage candidate preserves
reversibility of the Markov chain with respect to the target
distribution of interest (the posterior, in a Bayesian setting).
This delaying
rejection mechanism 
can be iterated for a fixed or random number of stages.

The higher stage proposal distributions can be designed in a very flexible way (using our intuition on the target at hand) and
are allowed to depend not only on the current position of the Markov chain but also
on the candidates
so far proposed and rejected (within each sweep). 
In some sense we can learn from our earlier mistakes. 
But notice that this form of local adaptation does not destroy the Markovian property since,
as soon as a candidate move is accepted, the rejected values are disregarded.
Thus DR allows partial local
adaptation of the proposal within each time step of the Markov
chain still retaining reversibility and Markovianity.
The advantage of DR over alternative ways of combining different
MH proposals or kernels, such as mixing and cycling \citep{tie94},
is that a hierarchy between kernels can be exploited so that kernels that are
easier to compute (in terms of CPU time) are tried first, thus saving
in terms of simulation time. Or moves that are more ``bold'' (bigger
variance of the proposal, for example) are tried at earlier stages
thus allowing the sampler to explore the state space more efficiently
following a sort of ``first bold'' versus ``second timid'' tennis-service
strategy.

Suppose the current position of the Markov chain is $X_t = \theta$. 
As in a regular MH, a candidate move $\theta_1$
is generated from a proposal $h_1(\theta,\cdot)$ and accepted with 
probability
\begin{equation}
\label{accept1}
\begin{split}
\alpha_1(\theta, \theta_1)  &= 1 \wedge \frac{p(\theta_1,y) h_1(\theta_1| \theta)}{p(\theta,y) h_1(\theta | \theta_1)} = 1 \wedge \frac{N_1}{D_1}. \\
\end{split}
\end{equation}
Note that the subscript in $h_1$ and $\alpha_1$ indicate that this is the first stage proposal and acceptance probability.
Upon rejection, instead of retaining the same position, $X_{t+1} = \theta$,
as we would do in a standard MH, a second stage move $\theta_2$ is
generated from a proposal distribution that is allowed to depend, not only on
the current position of the chain, but also on what we have just
proposed and rejected: $h_2( \theta_2 | \theta, \theta_1)$. The second stage
acceptance  probability is:
\begin{equation}
\tiny{
\label{a22}
\begin{split}
\alpha_2(\theta, \theta_1, \theta_2) & = 1 \wedge \frac{p(\theta_2,y) 
h_1(\theta_1 |\theta_2) h_2(\theta | \theta_2, \theta_1)[1 - \alpha_1(\theta_2, \theta_1)]}{p(\theta,y)
h_1(\theta_1 | \theta) h_2(\theta_2 | \theta, \theta_1)[1 - \alpha_1(\theta, \theta_1)]}  \\
&=1 \wedge \frac{N_2}{D_2}.\\
\end{split}}
\end{equation}
This process of delaying rejection can be iterated and the $i$-th
stage acceptance probability is, following \cite{mir01}:
\begin{equation}
\tiny{
\label{finalaccept}
\begin{split}
   \alpha_i & (\theta, \theta_1 \cdots \theta_i)  = \\
  & = 1 \wedge \frac{N_i}{D_i} \\
  & = 1 \wedge \left\{
    \frac{p(\theta_i,y) h_1(\theta_{i-1} | \theta_i) h_2( \theta_{i-2} | \theta_i, \theta_{i-1})
      \cdots h_i(\theta | \theta_i, \theta_{i-1} \cdots \theta_1)} {p(\theta,y) h_1(\theta_1 | \theta)
      h_2(\theta_2 | \theta, \theta_1) \cdots
      h_i(\theta_i | \theta, \theta_1 \cdots  \theta_{i-1})} \right. \\
  & \left. \frac{[1 - \alpha_1(\theta_i, \theta_{i-1})] [1 - \alpha_2(\theta_i, \theta_{i-1},
    \theta_{i-2})] \cdots [1 - \alpha_{i-1}(\theta_i, \cdots ,\theta_1)]}{[1 - \alpha_1(\theta, \theta_1)] [1 - \alpha_2(\theta, \theta_1, \theta_2)] \cdots [1 -
    \alpha_{i-1}(\theta, \theta_1, \cdots ,\theta_{i-1})]} \right\} \\
\end{split}
}
\end{equation}
\noindent If the $i$-th stage is reached, it means that $N_j < D_j$
for $j = 1, \cdots, i-1$, therefore $\alpha_{j}(\theta, \theta_1 \cdots \theta_{j})$ is simply $N_{j} / D_{j}, \; j = 1, \cdots, i-1$ and 
a recursive formula can be obtained:
$D_i = h_i(\theta \cdots \theta_i)(D_{i-1} - N_{i-1})$
which leads to:
\begin{equation} 
\begin{split}
    D_i &= h_i(\theta_i | \theta \cdots  ) [h_{i-1}(\theta_{i-1} | \theta \cdots )
    [h_{i-2}(\theta_{i-2} | \theta \cdots ) \cdots  \\
    & \;\;\; [h_2(\theta_2, | \theta, \theta_1) [ h_1(\theta_1| \theta) p(\theta,y) - N_1] - N_2]    - N_3] \cdots \\
    &  \;\;\;- N_{i-1}].
\end{split}
\end{equation}

Since 
reversibility
with respect to $p$ is preserved separately at each stage, the
process of delaying rejection can be interrupted at any stage.
The user can either  decide, in advance, to try at most, a fixed number of times to move away
from the current position or,
alternatively, upon each rejection,  toss a $\pi$-coin (i.e. a
coin with head probability equal to $\pi$), and if the outcome is head
move to a higher stage proposal, otherwise stay put.

 \cite{tie:mir99} prove that the DR strategy provides
MCMC estimators with smaller asymptotic variance than standard MH. This better performance holds no matter what is the function 
$f$ whose expectation relative to the target posterior  we want to
estimate (provided $f$ is squared integrable with respect to the target).
The performance of the approach has to be evaluated by weighting the improved asymptotic variance against the increased computational cost of the delayed rejection approach.

\section{Approximate Exchange Algorithm with Delayed Rejection (AEA+DR)}
\label{sec:exchangeDR}

The idea is to combine the DR strategy with the approximate exchange algorithm. We name this new algorithm the AEA+DR and different instances of it will be specified in 
subsequent sections depending of the (adaptive) proposal distribution used.
For the AEA+DR algorithm a theoretical modification of the $i$-th stage acceptance probability  is needed to take into account the fact that the target normalising constant depends on the parameter of interest. This is a novel methodological contribution that gives rise to an efficient MCMC sampler that can be used in general for doubly intractable problems. Efficiency is measure by the asymptotic variance of the resulting estimators. Indeed the AEA+DR dominates, 
in the Peskun sense \cite{pes73}, the original AEA in that the probability of moving away from the current position is higher.
Indeed, the intuition behind Peskun ordering is that, every time a Markov chain, used for MCMC purposes, retains the same position, it fails to explore the state space and the autocorrelation along its path increases, leading to a larger asymptotic variance of the sample path ergodic averages (the MCMC estimators). In a Metropolis-Hastings type algorithm the Markov chain stays put every time a candidate move is rejected. Thus, upon rejection, instead of advancing simulation time and retaining the same position, a second stage move is proposed. This attempt, by itself, increases the probability of moving away from the current position and thus the resulting AEA+DR algorithm achieves higher efficiency as measured by the effective sample size. Since the mechanism of delaying rejection is time consuming, a fair comparison should be made taking simulation time into account and thus considering the performance defined as ESS divided by simulation time. 

The first stage acceptance probability is unchanged relative to the standard AEA, and (recalling \eqref{eqn:acc_prob}) is given by:
\begin{equation}
\label{first:EXDR}
\begin{split}
\alpha_1 & (\theta, \theta_1)  =\\
& 1 \wedge
\frac
{ 
q(y|\theta_1) \;
p(\theta_1) \;
h_1(\theta|\theta_1) \;
q(y_1|\theta) \;
}
{ 
q(y|\theta) \;
p(\theta) \;
h_1(\theta_1|\theta) \;
q(y_1|\theta_1)\;
} 
\end{split}
\end{equation}

The second stage acceptance probability that preserves the detailed balance condition is:

\begin{equation}
\tiny{
\label{second:EXDR}
\begin{split}
\alpha_2 & (\theta, \theta_1, \theta_2)  =\\
& 1 \wedge
\frac
{ 
q(y|\theta_2) \;
p(\theta_2) \;
h_1(\theta_1|\theta_2) \;
h_2(\theta|\theta_2, \theta_1) \;
q(y_2|\theta) \;
\left[ 
      1- 
               \alpha_1(\theta_2,\theta_1)
\right]
}
{ 
q(y|\theta) \;
p(\theta) \;
h_1(\theta_1|\theta) \;
h_2(\theta_2|\theta, \theta_1) \;
q(y_2|\theta_2)\;
\left[ 
      1- 
               \alpha_1(\theta,\theta_1)
\right]
} 
\end{split}
}
\end{equation}
where $y_2$ are auxiliary data generated from the distribution $p(\cdot |\theta_2)$ which is the same likelihood distribution from which the observed data $y$ are assumed to have been sampled from. Higher stage acceptance probabilities are modified accordingly.
The second stage proposal of the delayed rejection version of the adaptive direction sampler (named ADS+DR) is designed to be negatively correlated with the first stage proposal following the idea of antithetic second stage  suggested in \cite{bed:dou:mou10}.

\section{Adaptive Approximate Exchange Algorithm (AAEA)}
\label{adaptation}

Three forms of adaptation of the Metropolis-Hastings proposal distribution (alternative to the ADS-AEA) are considered: vertical, horizontal and rectangular.
At simulation time $t$ there is a rectangular $t \times H$ family of particles  available: $\theta_{i,j}, i = 1, \cdots, t; j = 1, \cdots, H$.
Suppose we are interested in updating the position of particle $\theta_{t,h}$ (in the previous formulas this particle was simply indicated as $\theta$ with no subscripts).
To this aim  a random walk Metropolis-Hastings proposal is designed by taking a Gaussian distribution with mean equal to $\theta_{t,h}$ and variance-covariance matrix, 
given by the empirical variance (multiplied by $2.38^2 / d$ where $d$ is the model dimension, following \cite{rob:ros09}) of either: 
\begin{changemargin}{1cm}{0in}
\begin{itemize}
\item[\small AAEA-1]
all past particles along the same chain $h$ (vertical adaptation): $\theta_{i,h}, i = 1, \cdots, t-1$;
\item[\small AAEA-2]
all particles at the current time $t$ for all chains (horizontal adaptation): $\theta_{t,i}, i = 1, \cdots, H$;
\item[\small AAEA-3]
particles from all chains and all past simulations \citep[rectangular adaptation, aka inter-chain adaptation from][]{cra:ros:yan09}:  $\theta_{i,j}, i = 1, \cdots, t-1; j = 1, \cdots, H$.
\end{itemize}
\end{changemargin}
The sample covariance matrix used in AAEA-1 and AAEA-3 is computed recursively following the Equation (3) in \cite{haa:sak:tam01}.
\cite{rob:ros07} provide two conditions 
(``Containment'' and ``Diminishing Adaptation'')
that guarantee that a generic adaptive MCMC  is ergodic with respect to the  pro\-per stationary distribution.
In order to meet these conditions we follow the algorithm suggested in \cite{rob:ros07}
that,
with a small probability $\beta$ (set equal to $0.01$ in our simulation study), instead of using the adaptive proposal described above, uses the following static proposal: 
Normal distribution with variance-covariance matrix equal to $0.0025 I_d$ where $I_d$ is the identity matrix of size $d$, the model dimension.

We note that, when designing adaptive algorithms, it is usually not difficult to ensure directly that "Diminishing Adaptation" holds,
since adaption is user controlled one can always adapt less and less as the algorithm proceeds.
However, ``Containment''  is a  technical condition that avoids ``escape to infinity''. It always holds, for example, 
if the target has sub-exponential tails, or
if the state space is finite or
compact. The latter requirement is easily met by using a prior with compact support which is more than justified in the context of ERGM given the interpretation of the parameters.
The ``Containment''   condition, in general settings, may be more challenging to check. A careful review of sufficient conditions that ensure it
can be found in \cite{bai:rob:ros09}. We note that the horizontal adaptation scheme described above does not destroy the Markovian property since the covariance matrix is updated only based on information available at the current simulation time.

More sophisticated forms of MCMC adaptive  strategies could be used in the context of doubly intractable targets. We have not explored them further since, in the context of ERGMs, the adaptation procedures used, despite being simple are quite effective. We refer the interested reader to the tutorial by \cite{and:tho08} for a general 
framework of stochastic approximation
which allows one to systematically optimise generally used
criteria for designing adaptive MCMC algorithms, such as targeting a user specified acceptance probability.

As also discussed in \cite{cra:ros:yan09}, a question of interest in adaptive MCMC is whether one should wait a short or a long time before starting the adaptation.
Based on our simulation experience we found that the most effective strategy is to use the ADS approach during the burn-in phase 
and then switch to one of the adaptive algorithm mentioned above.
Furthermore, the simulation results presented use intensive adaptation \citep{gio:koh10} 
i.e. adaptation is performed at every iteration. 

We believe that more sophisticated forms of adaptations such as regional or tempered adaptation \citep{cra:ros:yan09}
are not needed in our setting. 
\section{Adaptive Approximate Exchange Algorithm with Delayed Rejection (AAEA+DR)}

The three adaptation schemes outlined in the previous section could be combined within the DR mechanism. 
For example at  first stage horizontal adaptation can be used since the resulting proposal is typically less computationally intensive to obtain 
(because $H$ is usually smaller than $t$ especially after the burn-in phase), at second stage we can try 
vertical adaptation and resort to rectangular adaptation only at third stage. The intuition behind this combination  is to use simple proposals first and resort to more 
refined proposals (typically more computationally intensive to construct, as rectangular adaptation)  only if really needed.

In the examples considered we follow a different rationale when combining the adaptive approximate exchange algorithm with the delayed rejection, namely, 
the second stage proposal is equal to the first stage one with the variance-covariance matrix rescaled by a factor of 0.5. 
In other words a more timid move is attempted at second stage. This is a very na\"ive form of delayed rejection but it is often quite effective (see for example \cite{gre:mir01,haa:lai:mir:sak06}).

\section{Examples}
\label{example}

\subsection{Introduction to the Results}
In this section we compare the adaptive direction sampler (ADS-AEA)  with three alternative forms of adaptation as defined in Section~\ref{adaptation}.

For each one of the four adaptive algorithms considered we have also implemented the corresponding two stage DR version as explained in Section~\ref{sec:exchangeDR}.
For vertical, horizontal and rectangular adaptation this is done by simply adding a second stage proposal which is identical to the first stage one except that the variance-covariance matrix is multiplied by $0.5$.\\
For the DR version of the adaptive direction sampler, upon rejection of the first stage candidate move 
$\theta_1^h = \theta^h + \gamma_1 \left( \theta_1^{h_1} - \theta_1^{h_2}\right) + \epsilon_1$,
the  second stage proposal is deterministically obtained from the first one by simply
going in the opposite direction relative to the current position:
$\theta_2^h = \theta^h - \gamma_1 \left( \theta^{h_1} - \theta^{h_2}\right) + \epsilon_1$ or, in other terms, $\theta^h_2 = 2 \theta^h_0 - \theta^h_1$.
This strategy  follows the idea of second stage antithetic proposal which has been proved in
\citep{bed:dou:mou10} to be generally highly effective.

We thus have a total of 8 different algorithms under comparison.
To our surprise the horizontal adaptation approach AAEA-2 outperforms all other adaptive algorithms (we were expecting rectangular adaptation to have a better performance given that more particles are used to learn the variance-covariance structure of the target distribution that is then used in the random walk Metropolis-Hastings sampler).
The performance of the horizontal adaptive algorithms is then further enhanced when combined with the delayed rejection strategy.

The DR version of each algorithm is always better than the corresponding single stage proposal in terms of ESS.
This simply confirms the fact that the DR dominates the corresponding single stage Metropolis-Hastings sampler in the Peskun ordering i.e. in terms of asymptotic variance of the resulting MCMC estimators (for a given number of sweeps). If the additional simulation time of the DR is also taken into account in the comparison (i.e. if we consider the ESS per unit computational time), the DR still outperforms the corresponding single stage proposal in all cases but for the adaptive direction sampler (this is because the second stage proposal of the 
adaptive direction sampler, ADS+DR, has a very small acceptance probability due to the structure of the first stage proposal and the antithetic move implemented at second stage).

In the next 3 subsections we present 3 examples of increasing complexity.
The Florentine Marriage Network has 16 nodes and the proposed ERGM has 3 parameters; the Karate Club network has 34 nodes and 3 parameters; while the Faux Mesa High School Network has 208 nodes and  9 parameters.
We anticipate that, as the dimensions and the complexity of the model increase, the efficiency of the 3 proposed adaptive strategies with delayed rejection (that turn out to be the winning strategies) becomes more and more comparable in terms performance while, in terms of ESS (a more neutral measure in that it is not affected by coding ability), the best algorithm is rectangular adaptation with delayed rejection.

Horizontal adaptation performs better because calculating the covariance of a small set of points is cheaper than 
calculating the covariance matrix in vertical and rectangular adaptation where more points enter in the estimation of the covariance.
Indeed the CPU time needed to run the 3 adaptation scheme follows, in general, this ordering CPU-horizontal $<$ CPY-vertical $<$ CPU-rectangular
and this is because, when computing the variance-covariance matrix, there and increasingly more particles entering the computation for vertical versus horizontal  and for rectangular versus vertical. Furthermore, the number of particles used in vertical and rectangular adaptation increases with simulation length while it is a static value (equal to the number of chains) for horizontal adaptation. Of course the intuition is that more particles lead to a better estimate and thus more efficient sampler resulting in larger ESS. There is thus a trade-off that leads to a competitive advantage of horizontal adaptation in small and simple models. This advantage washes out for larger and more complex models.

Horizontal adaptation performs better for targets of small dimension, because calculating the covariance matrix of a small set of points turns out to be cheaper (in terms of CPU time) than computing the covariance  in vertical and rectangular adaptation. There is, of course, an interplay between the number of chains and the number of iterations used. 
In horizontal adaptation, the number of chains has to be ``large enough'' in order to have the population points spread about the target distribution. 
As a rule of thumb, in horizontal adaptation, the number of chains should grow with the square of the dimension of the target since we need to estimate its $d$ dimensional covariance matrix. Thus we conjecture that the better performance of horizontal adaptation will wash out as $d$ grows (more network statistics)
and the computational complexity of the problem increases (more nodes), indeed our simulation results agree with this intuition (see results in the last example). 

\subsection{Florentine Marriage Network}
\label{sec:flo}
Let us consider again the Florentine marriage network and model defined in Equation~\ref{eqn:model1}.  
We use a total number of $24,000$ iterations for estimating the posterior density and 4 different approaches: 
\begin{itemize}
\item ADS-AEA consists of 6 chains of $4,000$ iterations each;
\item AAEA-1 (vertical adaptation) consists of 6 chains of $4,000$ iterations each;
\item AAEA-2 (horizontal adaptation) consisting of 24 chains of $1,000$ iterations each;
\item AAEA-3 (rectangular adaptation) consisting of 6 chains of $4,000$ iterations each.
\end{itemize}

In Table~\ref{tab:flo2} are displayed the posterior parameter estimates and effective sample size calculated for the AAEA-2 and AAEA-2+DR which turned out to be the best approaches in terms of performance. In particular the AAEA-2+DR yields a variance reduction of about 83\% compared to the ADS-AEA.

\begin{table}
\caption{Florentine marriage network - Posterior parameter estimates and effective sample size (ESS) for model~\ref{eqn:model1}.}
\label{tab:flo2}
\begin{tabular}{l|ccc}
\hline\noalign{\smallskip}
       & \multicolumn{3}{c}{AAEA-2 (horizontal adaptation)} \\
\noalign{\smallskip}\hline\noalign{\smallskip}
       & $\theta^{(1)}$ (edges) & $\theta^{(2)}$ (2-stars) & $\theta^{(3)}$ (3-stars)\\
\noalign{\smallskip}\hline\noalign{\smallskip}
Post. mean & -1.47 & 0.05 & -0.06\\
Post. sd   & 1.86 & 0.69 & 0.36\\
\noalign{\smallskip}\hline
\hline\noalign{\smallskip}
       & \multicolumn{3}{c}{AAEA-2+DR (horizontal adaptation + DR)} \\
\noalign{\smallskip}\hline\noalign{\smallskip}
       & $\theta^{(1)}$ (edges) & $\theta^{(2)}$ (2-stars) & $\theta^{(3)}$ (3-stars)\\
\noalign{\smallskip}\hline\noalign{\smallskip}
Post. mean & -1.61 & 0.08 & -0.06\\
Post. sd   & 1.55 & 0.53 & 0.25\\
\noalign{\smallskip}\hline
\end{tabular}
\end{table}

\begin{figure}[htp]
\includegraphics[scale=.7]{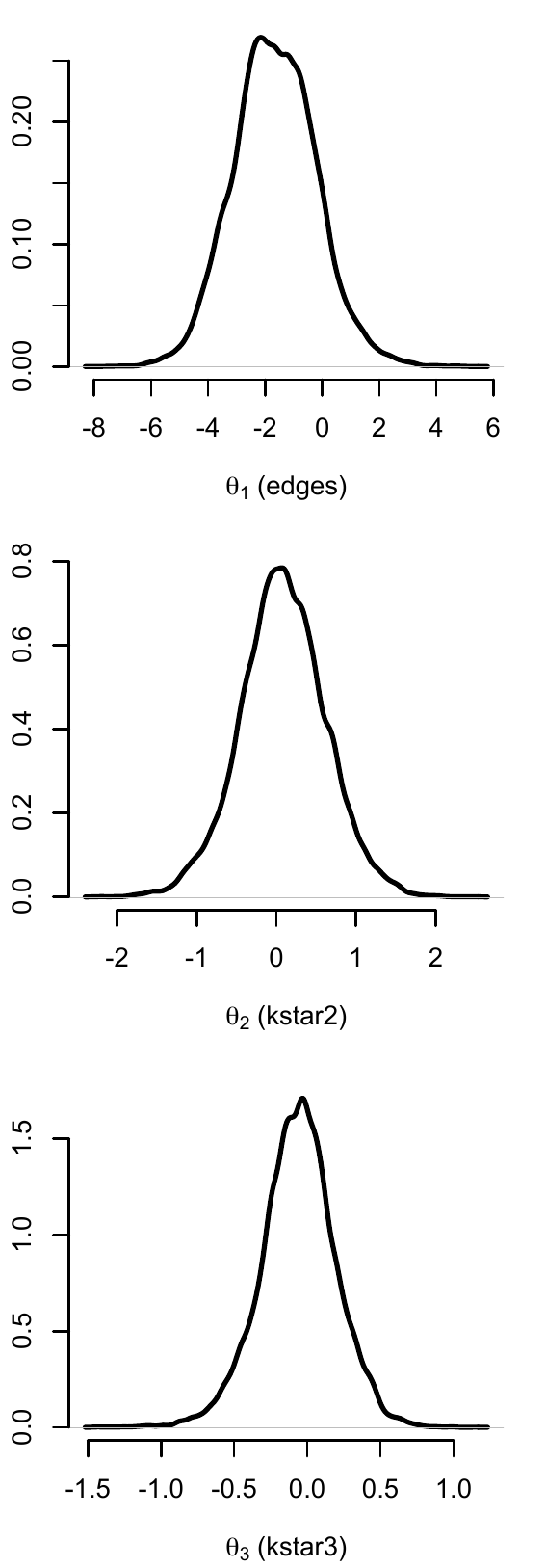}
\includegraphics[scale=.7]{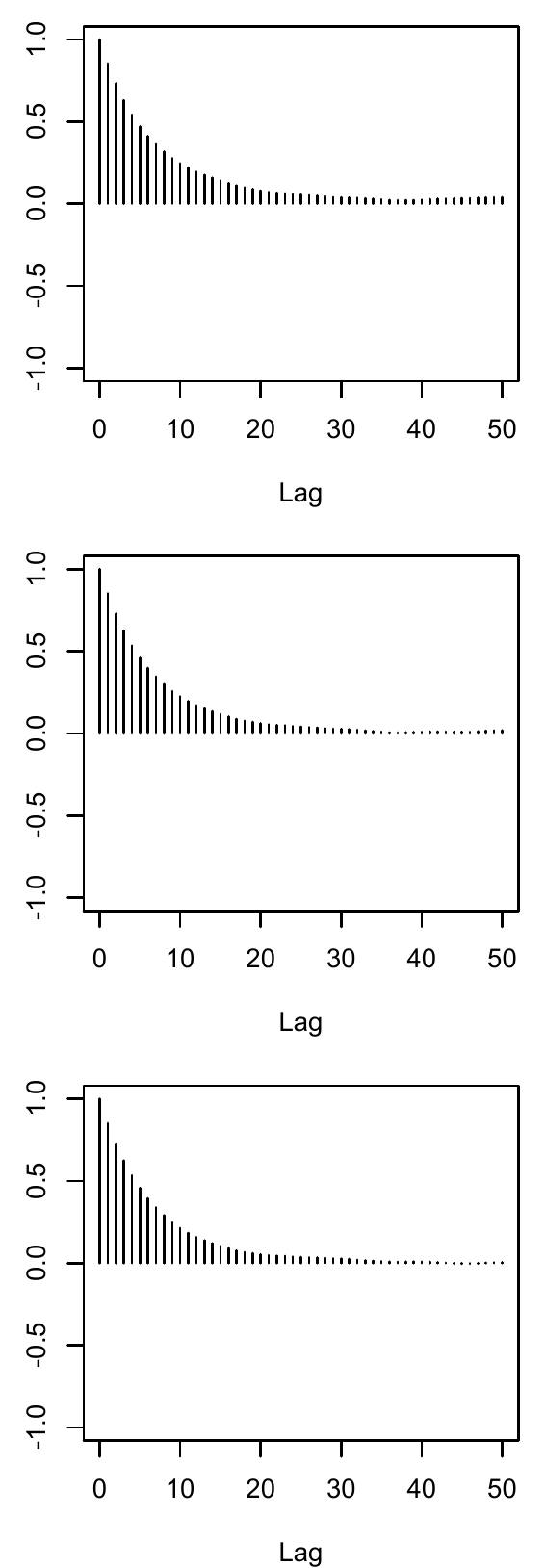}
\caption{Florentine marriage network - MCMC diagnostics for the AAEA-2+DR.}
\label{fig:flo5}
\end{figure}

In Figure~\ref{fig:flo5} it can be seen that the autocorrelations of the parameter estimates returned by the AAEA-2+DR decay quicker than the autocorrelations of the other two approaches displayed in Figures~\ref{fig:flo0}.
The AAEA-2 algorithm outperforms the ADS-AEA in terms of both ESS (20\%) and performance (20\%). The AAEA-2+DR outperforms the AAEA-2 in terms of ESS of about 60\% and performance of about 15\% (Table~\ref{tab:flo_ess}). Computing times can be calculated as ESS / Performance.

\begin{table*}
\caption{Florentine marriage network -  effective sample size (ESS) and performance for each algorithm for model~\ref{eqn:model1} based on 100 simulations.}
\label{tab:flo_ess}
\begin{tabular}{l|c|c|c|c}
\hline\noalign{\smallskip}
       & ADS-AEA & AAEA-1 & AAEA-2 & AAEA-3 \\
\noalign{\smallskip}\hline\noalign{\smallskip}
ESS              & 755 & 753 & 896 & 833\\
Performance (per sec) & 33 & 27 & 38 & 28\\
\noalign{\smallskip}\hline
\hline\noalign{\smallskip}
       & ADS-AEA+DR & AAEA-1+DR & AAEA-2+DR & AAEA-3+DR \\
\noalign{\smallskip}\hline\noalign{\smallskip}
ESS & 771 & 1478 & 1385 & 1201\\
Performance (per sec) & 33 & 33 & 41 & 34\\
\noalign{\smallskip}\hline
\hline\noalign{\smallskip}
\end{tabular}
\end{table*}

In Table~\ref{corr:flo} it is possible to observe the correlation matrix between the parameters in the posterior distribution. There is a very strong negative correlation between all the parameters of the model.

\begin{table}[htp]
\caption{Florentine marriage network - Posterior correlation matrix between the parameters in the distribution for model~\ref{eqn:model1}.}
\label{corr:flo}
\centering
\begin{tabular}{r|rrr}
\hline\noalign{\smallskip}
       & $\theta^{(1)}$ & $\theta^{(2)}$ & $\theta^{(3)}$\\
\noalign{\smallskip}\hline\noalign{\smallskip}
  $\theta^{(1)}$ & 1.00 & -0.94 & -0.80 \\ 
  $\theta^{(2)}$ & . & 1.00 & -0.94 \\ 
  $\theta^{(3)}$ & . & . & 1.00 \\ 
\noalign{\smallskip}\hline
\end{tabular}
\end{table}


\subsection{Karate club network}
\label{sec:karate}
\begin{figure}[htp]
\centering
\includegraphics[scale=.65]{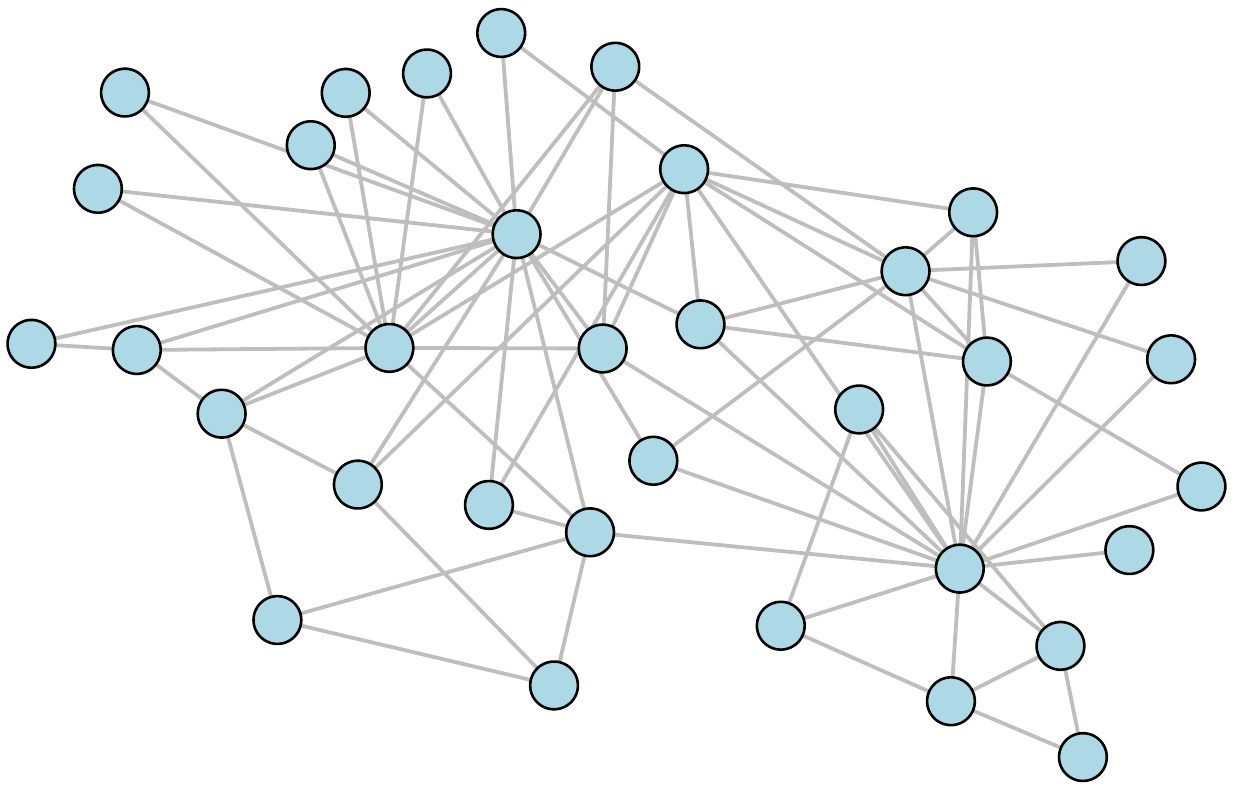}
\caption{Zachary karate club network graph.}
\label{fig:zach-graph}
\end{figure}

This example concerns the karate club network \citep{zac77} displayed in Figure~\ref{fig:zach-graph} which represents friendship relations between 34 members of a karate club at a US university in the 1970. 

We propose to estimate the following 3-dimensional model using the network statistics proposed by \cite{sni:pat:rob:han06}:
\begin{equation}
q(y|\theta) = \exp \left\lbrace 
\theta^{(1)} s_1(y) + 
\theta^{(2)} v(y,\phi_u) + 
\theta^{(3)} u(y,\phi_v) 
\right\rbrace
\label{eqn:model2}
\end{equation}
where
\begin{center}
\begin{tabular}{ll}
$s_{1}(y) = \sum_{i<j}y_{ij}$ number of edges\\[.1cm]
$v(y,\phi_v) = e^{\phi_v} \sum_{i=1}^{n-2}\left \{ 1-\left( 1 - e^{-\phi_v} \right)^{i} \right \} EP_i(y)$ & \\
\quad geometrically weighted edgewise shared partners\\ \quad (GWESP)\\[.1cm]
$u(y,\phi_u) = 
e^{\phi_u} \sum_{i=1}^{n-1} 
\left\{ 1- \left( 1 - e^{-\phi_u} \right )^{i} \right \} D_i(y)$ & \\
\quad geometrically weighted degrees (GWD)\\
\end{tabular}
\end{center}
where $EP_i(y)$ and $D_i(y)$ are the edgewise shared partners and degree distributions respectively. We set $\phi_{u} = \phi_{v} = \log(2)$ so that the model is a non-curved ERGM \citep{hun:han06}.
The prior setting is the same as the one in Section~\ref{sec:flo1}: $p(\theta)\sim \mathcal{N}(0,100 I_3)$. The tuning parameters for the ADS proposal are: $\gamma = 0.9$ and $\epsilon \sim \mathcal{N}(0,0.0025 I_d)$ so that the overall acceptance rate is around $21\%$. The auxiliary chain consists of $100$ iterations and a total number of $24,000$ main iterations is used. The number of chains used in the various strategies is the same as in the previous example in Section~\ref{sec:flo}.

In this example, as happened in the teenage friendship network above, the AAEA-3 outperforms the AAEA-2 in terms of variance reduction of about  40\% but not in terms of performance. For this reason AAEA-2 is still to be preferred.

\begin{figure}[htp]
\includegraphics[scale=.7]{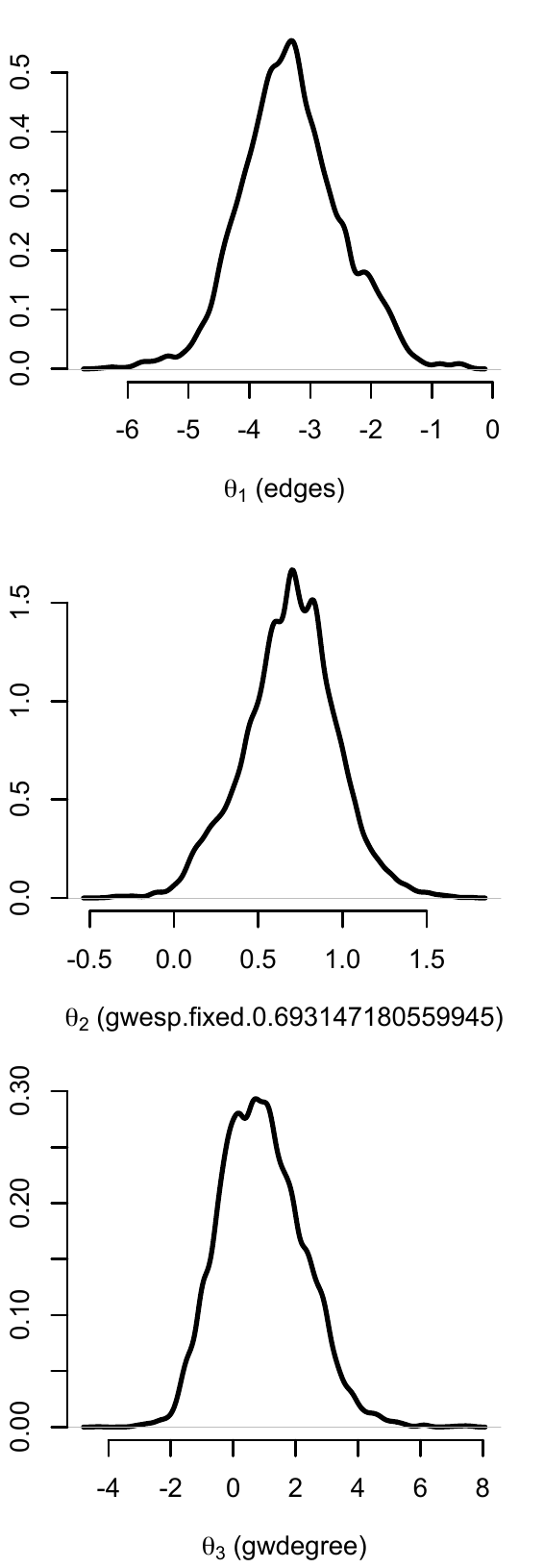}
\includegraphics[scale=.7]{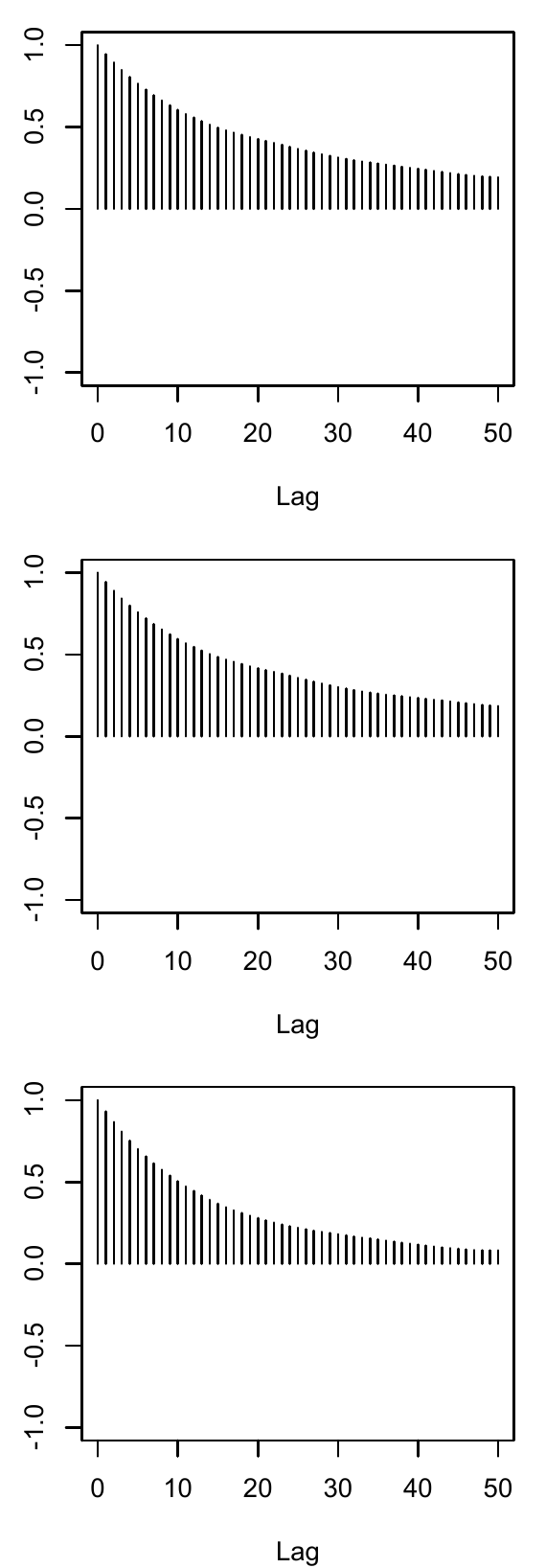}
\caption{Zachary karate club network - MCMC diagnostics for the ADS-AEA.}
\label{fig:zach0}
\end{figure}

\begin{figure}[htp]
\includegraphics[scale=.7]{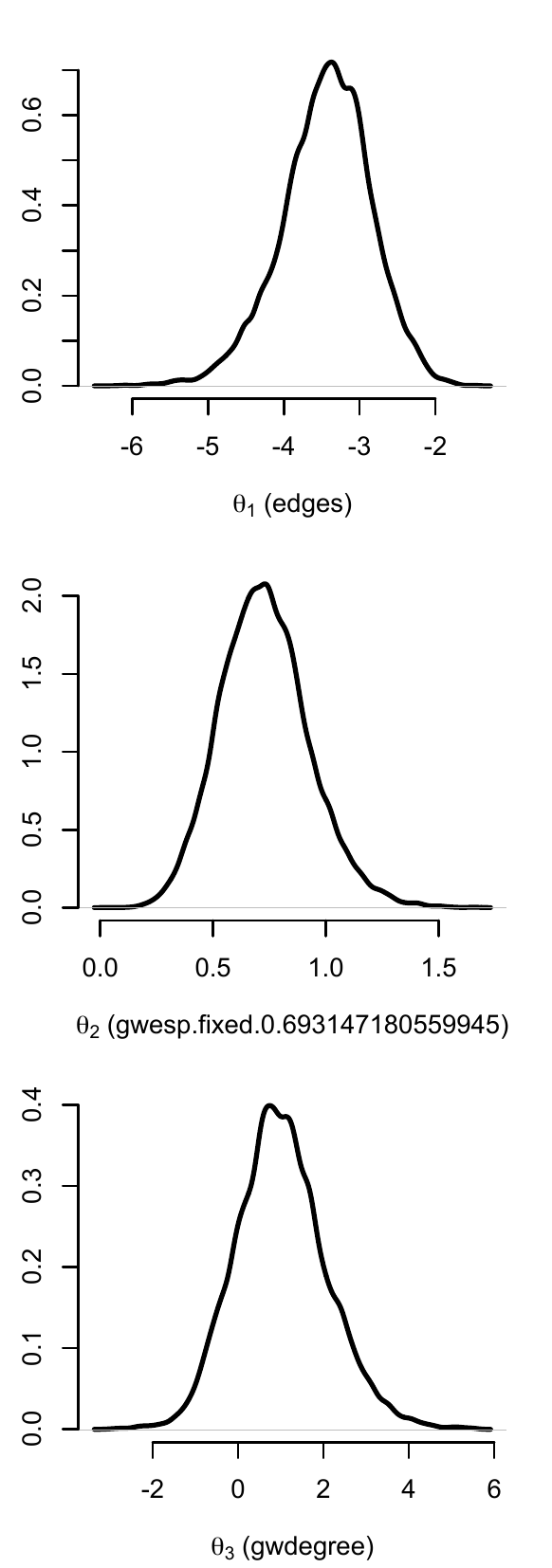}
\includegraphics[scale=.7]{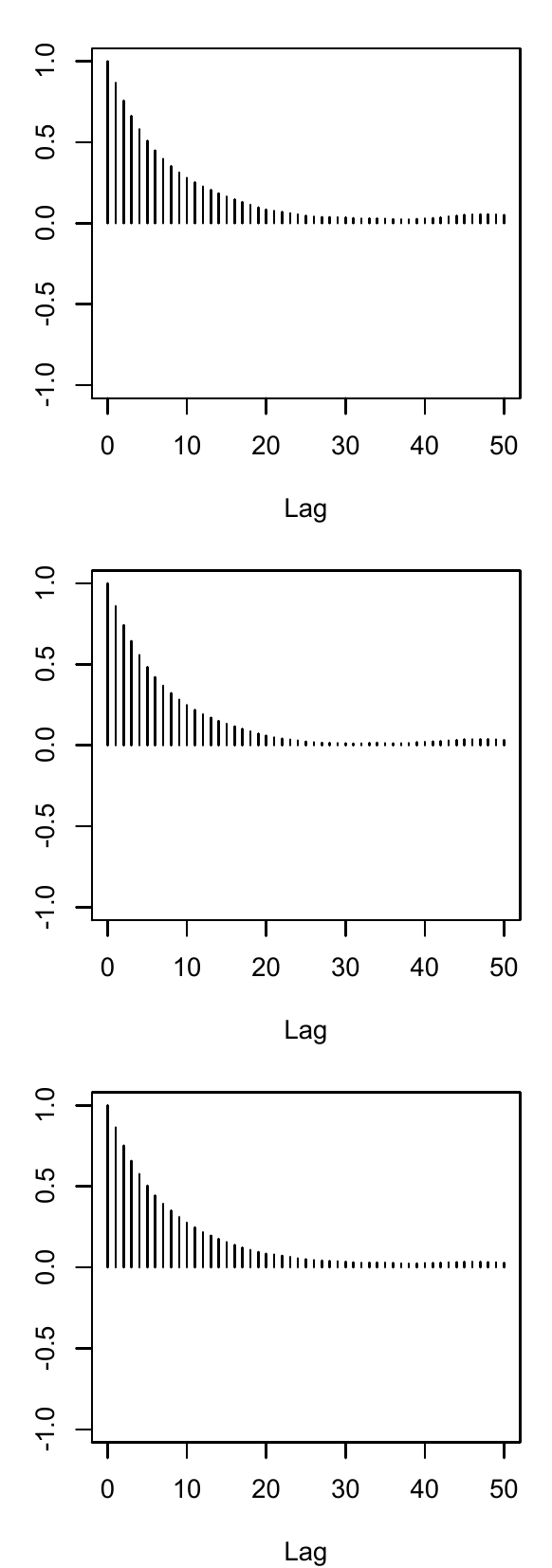}
\caption{Zachary karate club network - MCMC diagnostics for the AAEA-2+DR.}
\label{fig:zach5}
\end{figure}

In Figure~\ref{fig:zach5} it can be seen that the autocorrelations of the parameters for the AAEA-2 approach decay quicker than the autocorrelations given by the other methods as shown in Figure~\ref{fig:zach0}.
The AAEA-2 outperforms the ADS-AEA of about 12\% in terms of performance whereas the AAEA-2+DR makes a further improvement of about 20\% with respect to the AAEA-2+DR (see Table~\ref{tab:zach_ess}).

\begin{table}
\caption{Zachary karate club network - Posterior parameter estimates
for model~\ref{eqn:model2}.}
\label{tab:zach}
\begin{tabular}{l|cccc}
\hline\noalign{\smallskip}
       & \multicolumn{3}{c}{ADS-AEA} \\
\noalign{\smallskip}\hline\noalign{\smallskip}
       & $\theta^{(1)}$ (edges) & $\theta^{(2)}$ (gwesp) & $\theta^{(3)}$ (gwdegree)\\
\noalign{\smallskip}\hline\noalign{\smallskip}
Post. mean & -3.51 & 0.74 & 1.18\\
Post. sd   & 0.62 & 0.21 & 1.12\\
\noalign{\smallskip}\hline
\hline\noalign{\smallskip}
       & \multicolumn{3}{c}{AAEA-2+DR (horizontal adaptation + DR)} \\
\noalign{\smallskip}\hline\noalign{\smallskip}
       & $\theta^{(1)}$ (edges) & $\theta^{(2)}$ (gwesp) & $\theta^{(3)}$ (gwdegree)\\
\noalign{\smallskip}\hline\noalign{\smallskip}
Post. mean & -3.44 & 0.72 & 1.01\\
Post. sd   & 0.59 & 0.21 & 1.07\\
\noalign{\smallskip}\hline
\hline\noalign{\smallskip}
\end{tabular}
\end{table}

\begin{table*}
\caption{Zachary karate club network -  effective sample size (ESS) and performance for each algorithm for model~\ref{eqn:model2} based on 100 simulations.}
\label{tab:zach_ess}
\begin{tabular}{l|c|c|c|c}
\hline\noalign{\smallskip}
       & ADS-AEA & AAEA-1 & AAEA-2 & AAEA-3 \\
\noalign{\smallskip}\hline\noalign{\smallskip}
ESS & 840 & 724 & 605 & 776\\
Performance (per sec) & 21 & 23 & 23 & 22\\
\noalign{\smallskip}\hline
\hline\noalign{\smallskip}
       & ADS-AEA+DR & AAEA-1+DR & AAEA-2+DR & AAEA-3+DR \\
\noalign{\smallskip}\hline\noalign{\smallskip}
ESS & 850 & 1410 & 1306 & 1418\\
Performance (per sec) & 20 & 26 & 27 & 25\\
\noalign{\smallskip}\hline
\hline\noalign{\smallskip}
\end{tabular}
\end{table*}

As in the Florentine marriage network example, we can observe (Table~\ref{corr:zach}) that there is a strong negative posterior correlation between parameters $\theta^{(1)}$ and $\theta^{(2)}$ and between $\theta^{(1)}$ and $\theta^{(3)}$.

\begin{table}[htp]
\caption{Zachary karate club network - Posterior correlation matrix between the parameters in the distribution for model~\ref{eqn:model2}.}
\label{corr:zach}
\centering
\begin{tabular}{r|rrr}
\hline\noalign{\smallskip}
& $\theta^{(1)}$ & $\theta^{(2)}$ & $\theta^{(3)}$\\
\noalign{\smallskip}\hline\noalign{\smallskip}
$\theta^{(1)}$ & 1.00 & -0.80 & -0.75 \\ 
$\theta^{(2)}$ & . & 1.00 & 0.37 \\ 
$\theta^{(3)}$ & . & . & 1.00 \\ 
\noalign{\smallskip}\hline
\end{tabular}
\end{table}

Generally a strong correlation between parameters in the posterior distribution hampers the behaviour of vanilla 
MCMC schemes. In fact high posterior correlation can slow down the motion of the chain towards equilibrium distribution. It is in this case that the adaptive approximate exchange algorithm with delayed rejection (AAEA-2+DR) gives the best performance compared to the adaptive direction sampling approximate exchange algorithm.

\subsection{Faux Mesa High School Network}
\label{sec:school}
\begin{figure}[htp]
\centering
\includegraphics[scale=.675]{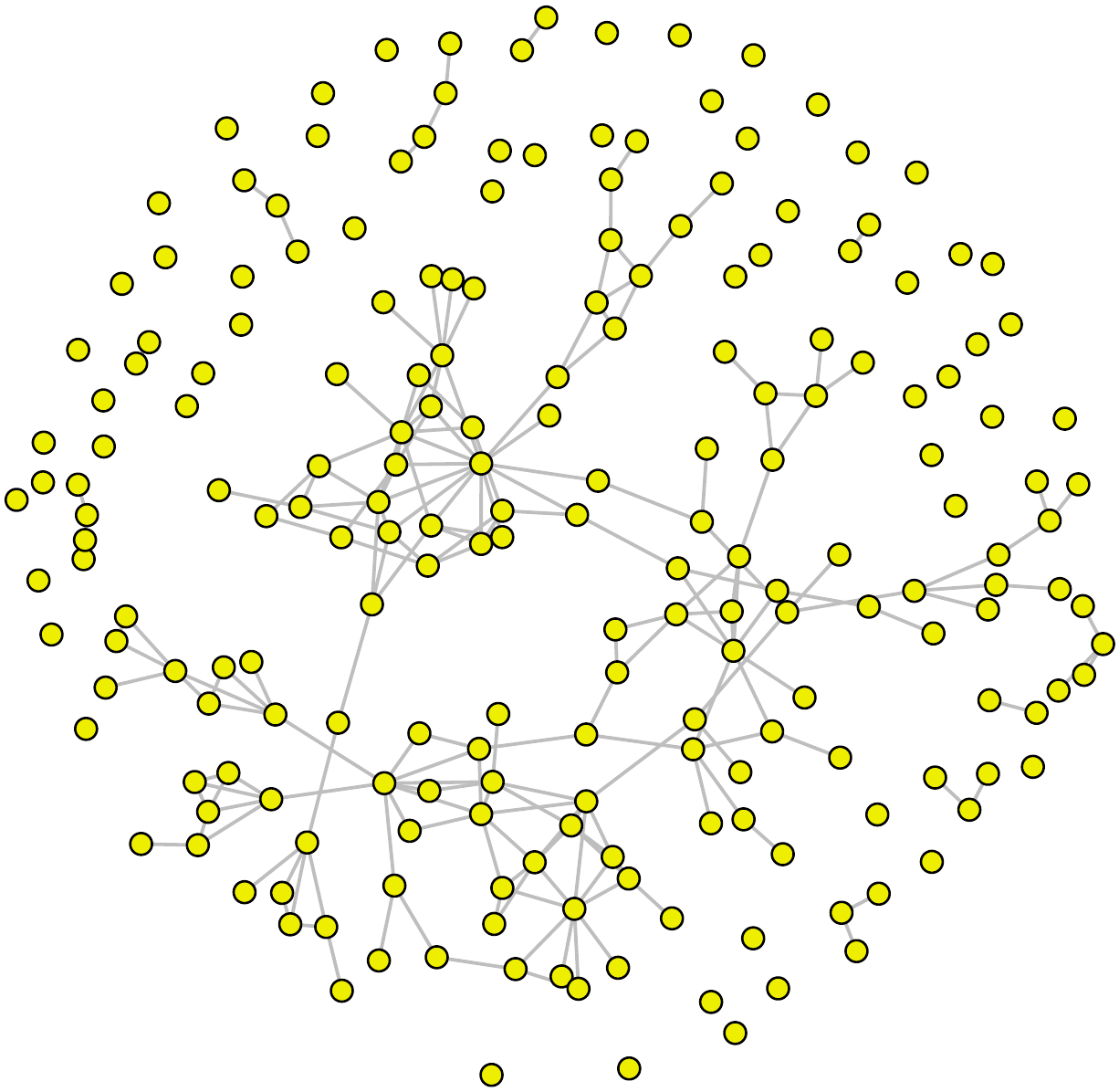}
\caption{Faux Mesa High School friendship network graph.}
\label{fig:school}
\end{figure}

In this example we revisit a well known network dataset (Figure~\ref{fig:school}) in social science concerning friendship relations in a school community of 203 students \cite{han:hun:but:goo:mor07}. The vertex attributes $x$ that we are interested in are ``grade'' (it takes values 7 through 12 indicating each student's grade in school) and ``sex'' of each student. 

The main focus is on the factor attribute effects (which give information about the tendency of a node with a specific attribute to form an edge in the network) and on the transitivity effect expressed by the GWESP and GWD statistics defined in Section \ref{sec:karate} with $\phi_{u} = \phi_{v} = 1$.
 
The model we propose to estimate is defined by the following 9 network statistics:
\begin{center}
\begin{tabular}{l}
$s_{1}(y) = \sum_{i<j}y_{ij}$ number of edges\\[.1cm]
$s_{2}(y,x) = \sum_{i<j}y_{ij} (\mathds{1}_{(grade_{i} = 8)} + \mathds{1}_{(grade_{j} = 8)})$\\
\qquad node factor for ``grade'' = 8\\[.1cm]
$s_{3}(y,x)= \sum_{i<j}y_{ij} (\mathds{1}_{(grade_{i} = 9)} + \mathds{1}_{(grade_{j} = 9)})$\\
\qquad node factor for ``grade'' = 9\\[.1cm]
$s_{4}(y,x)= \sum_{i<j}y_{ij} (\mathds{1}_{(grede_{i} = 10)} + \mathds{1}_{(grade_{j} = 10)})$\\
\qquad node factor for ``grade'' = 10\\[.1cm]
$s_{5}(y,x)= \sum_{i<j}y_{ij} (\mathds{1}_{(grade_{i} = 11)} + \mathds{1}_{(grade_{j} = 11)})$\\
\qquad node factor for ``grade'' = 11\\[.1cm]
$s_{6}(y,x)= \sum_{i<j}y_{ij} (\mathds{1}_{(grade_{i} = 12)} + \mathds{1}_{(grade_{j} = 12)})$\\
\qquad node factor for ``grade'' = 12\\[.1cm]
$s_{7}(y,x)= \sum_{i<j}y_{ij} (\mathds{1}_{(sex_{i} = M)} + \mathds{1}_{(sex_{j} = M)})$\\
\qquad node factor for ``sex = male"\\[.1cm]
$s_{8}(y) = v(y,\phi_v)$ GWESP\\[.1cm]
$s_{9}(y) = u(y,\phi_u)$ GWD\\
\end{tabular}
\end{center}
where $\mathds{1}_{(\cdot)}$ is the indicator function.

The tuning parameters for the ADS proposal $\gamma = 0.3$ and $\epsilon \sim \mathcal{N}(0,0.0025 I_d)$ are chosen so as to obtain the overall acceptance rate is around $21\%$.
$5,000$ auxiliary iterations 
are used for network simulation and $60,000$ main iterations are used for estimating the posterior density of model defined above: 
\begin{itemize}
\item ADS-AEA consists of 20 chains of $3,000$ iterations each;
\item AAEA-1 (vertical adaptation) consists of 30 chains of $2,000$ iterations each;
\item AAEA-2 (horizontal adaptation) consists of 20 chains of $3,000$ iterations each;
\item AAEA-3 (rectangular adaptation) consists of 20 chains of $3,000$ iterations each.
\end{itemize}

\begin{table*}
\caption{Faux Mesa High School network - Posterior parameter estimates.}
\label{tab:school}
\begin{tabular}{l|cccccccccc}
\hline\noalign{\smallskip}
       & \multicolumn{9}{c}{ADS-AEA} \\
\noalign{\smallskip}\hline\noalign{\smallskip}
       & $\theta^{(1)}$ & $\theta^{(2)}$ & $\theta^{(3)}$ & $\theta^{(4)}$ & $\theta^{(5)}$ & $\theta^{(6)}$ & $\theta^{(7)}$ & $\theta^{(8)}$ & $\theta^{(9)}$ \\
\noalign{\smallskip}\hline\noalign{\smallskip}
Post. mean & -5.53 & -0.15 & -0.09 & -0.04 &  -0.12 & 0.20 & -0.18 & 0.28 & 1.53\\
Post. sd       & 0.33 &   0.15 & 0.17 &  0.21  &  0.18 & 0.23 & 0.12 & 0.25 & 0.12\\
\noalign{\smallskip}\hline
\hline\noalign{\smallskip}
       & \multicolumn{9}{c}{AAEA-2+DR (horizontal adaptation + DR)} \\
\noalign{\smallskip}\hline\noalign{\smallskip}
       & $\theta^{(1)}$ & $\theta^{(2)}$ & $\theta^{(3)}$ & $\theta^{(4)}$ & $\theta^{(5)}$ & $\theta^{(6)}$ & $\theta^{(7)}$ & $\theta^{(8)}$ & $\theta^{(9)}$ \\
\noalign{\smallskip}\hline\noalign{\smallskip}
Post. mean & -5.48 & -0.14 & -0.09 & -0.04 & -0.11 & 0.19 & -0.17 & 0.27 & 1.52\\
Post. sd       &  0.30 &  0.12 & 0.13 & 0.19 & 0.16 & 0.20 &  0.10 & 0.23 & 0.11\\
\noalign{\smallskip}\hline
\hline\noalign{\smallskip}
\end{tabular}
\end{table*}

In Table~\ref{tab:school_ess}, the adaptive algorithms with delayed rejection outperform the ADS-AEA in terms of both variance reduction and performance. All the adaptive algorithms with delayed rejection deliver the same results in terms of performance.

\begin{table*}
\caption{Faux Mesa High School network - ESS and performance for each algorithm based on 10 simulations.}
\label{tab:school_ess}
\begin{tabular}{l|c|c|c|c}
\hline\noalign{\smallskip}
       & ADS-AEA & AAEA-1 & AAEA-2 & AAEA-3 \\
\noalign{\smallskip}\hline\noalign{\smallskip}
ESS & 667 & 1041 & 1008 & 1094\\
Performance (per sec) & 1.8 & 2.3 & 2.1 & 2.2\\
\noalign{\smallskip}\hline
\hline\noalign{\smallskip}
       & ADS-AEA+DR & AAEA-1+DR & AAEA-2+DR & AAEA-3+DR \\
\noalign{\smallskip}\hline\noalign{\smallskip}
ESS & 873 & 1376 & 1320 & 1440\\
Performance (per sec) & 1.4 & 2.6 & 2.6 & 2.6\\
\noalign{\smallskip}\hline
\hline\noalign{\smallskip}
\end{tabular}
\end{table*}

As in the previous examples, we can observe (Table~\ref{corr:school}) that there is a strong negative posterior correlation between parameters $\theta^{(1)}$ and $\theta^{(8)}$ and between $\theta^{(1)}$ and $\theta^{(9)}$.

\begin{table*}[htp]
\caption{Faux Mesa High School network - Posterior correlation matrix between the parameters.}
\label{corr:school}
\centering
\begin{tabular}{r|rrrrrrrrr}
\hline\noalign{\smallskip}
& $\theta^{(1)}$ & $\theta^{(2)}$ & $\theta^{(3)}$ & $\theta^{(4)}$ & $\theta^{(5)}$ & $\theta^{(6)}$ & $\theta^{(7)}$ & $\theta^{(8)}$ & $\theta^{(9)}$ \\
\noalign{\smallskip}\hline\noalign{\smallskip}
$\theta^{(1)}$ &  1.00 & -0.04 & -0.08 &  0.08 & -0.18 & -0.25 & -0.16 & -0.83 & -0.80\\ 
$\theta^{(2)}$ &  . & 1.00 & 0.32 & 0.34 & 0.13 & 0.17 & -0.10 & -0.20 & -0.13\\
$\theta^{(3)}$ &  . &  . & 1.00 & 0.23 & 0.15 & 0.23 & -0.08 & -0.21 & -0.05\\
$\theta^{(4)}$ &  . &  . &  . & 1.00 & -0.04 &  0.24 & -0.13 & -0.25 & -0.17\\ 
$\theta^{(5)}$ &  . &  . &  . &  . & 1.00 & 0.05 & -0.07 & 0.04 & 0.07\\
$\theta^{(6)}$ &  . &  . &  . &  . &  . &   1.00 &  0.03 & 0.01 &  0.08\\
$\theta^{(7)}$ &  . &  . &  . &  . &  . &  . &  1.00 & -0.07 & -0.08\\ 
$\theta^{(8)}$ &  . & . & . &  . &  . &  . &  . &  1.00 &  0.73\\
$\theta^{(9)}$ &  . & . & . & . & . &  . & . & . & 1.00\\
\noalign{\smallskip}\hline
\end{tabular}
\end{table*}

From the results displayed in Table~\ref{tab:school} we can conclude that the network is very sparse ($\theta^{(1)}$ negative) and that students having the same gender seem to create friendship connections ($\theta^{(7)}$ negative). The transitivity effect expressed by $\theta^{(8)}$ and the popularity effect expressed by $\theta^{(9)}$ are important features of the network.

\section{Conclusions}
\label{sec:conclusions}
	
The exchange algorithm of \cite{mur:gha:mac06}
makes the computation of the MH acceptance probability feasible even for target distributions
whose normalizing constant depends on the parameter of interest (doubly intractable problems).

The {\em approximate exchange algorithm}, due to \cite{cai:fri11}, modifies  the original exchange algorithm and makes it applicable also in settings where sampling from the assumed data generating process is not feasible. This is the case for exponential random graphs the model we focus on in this paper.

The {\em delayed rejection} strategy allows to locally adapt the proposal distribution within each sweep of a MH algorithm at the cost of additional computational time.

The {\em adaptive random walk proposal} of \cite{haa:sak:tam01} revised by \cite{rob:ros09} allows for global adaptation between MH iterations. This learning from the past process is also expensive from a computational point of view.
 
These three ingredients are combined in different ways within the approximate exchange algorithm (AEA) to avoid the computation of intractable normalising constant that appears in exponential random graph models. This gives rise to the AEA+DR: a new methodology to sample doubly intractable target distributions which achieves variance reduction relative to the adaptive direction sampling approximate exchange algorithm of \cite{cai:fri11} implemented in the {\sf Bergm} package for {\sf R} \citep{cai:fri14}, which is our benchmark.

The 8 algorithms under comparison (seven of which are original contributions) are tested on three examples.
Consistently, the best combination (in terms of ESS for fixed simulation time), is given by the {\em horizontal adaptive approximate exchange algorithm with delayed rejection}, which achieves a variance reduction that varies between 55\% and 98\% (relative to the benchmark). 

This translates into a better performance varying from 25\% to 40\%, if the extra simulation time, due to the delayed rejection mechanism and the adaptation procedure, is taken  into account. The strongest improvements are obtained in the examples with highly correlated posterior distributions.

The applicability of the proposed methodology goes beyond the social network context as it works for any doubly intractable target.

The delayed rejection strategy and the form of adaptation proposed in the present paper have been implemented in the {\sf Bergm} package.

\bibliographystyle{spbasic} 
\bibliography{BIBLIOGRAPHY}

\end{document}